\documentclass[aps,prl,preprint,amsfonts,amsmath,amssymb,floatfix,10pt,reprint]{revtex4-1}
\usepackage{braket} 
\usepackage{graphicx}
\usepackage{algorithm}
\usepackage{algpseudocode}
\floatname{algorithm}{Procedure}

\usepackage{hyperref}

\usepackage{color}
\usepackage[normalem]{ulem}

\graphicspath{{./}}

\algnewcommand{\LineComment}[1]{\State\(\triangleright\) #1}

\begin{document}
\title{An adaptive algorithm for quantum circuit simulation}

\author{Roman Schutski}
\affiliation{Center for Computational and Data-Intensive Science and Engineering, Skoltech, Skolkovo Innovation Center, Moscow Region, 121205, Russian Federation}

\author{Danil Lykov}
\affiliation{Moscow Institute of Physics and Technology, Dolgoprudny, Moscow Region, 141701, Russian Federation}

\author{Ivan Oseledets}
\affiliation{Center for Computational and Data-Intensive Science and Engineering, Skoltech, Skolkovo Innovation Center, Moscow Region, 121205, Russian Federation}
\date{\today}

\begin{abstract}
Efficient simulation of quantum computers is essential for the development and validation of near-term quantum devices and the research on quantum algorithms. Up to date, two main approaches to simulation were in use, based on either full state or single amplitude evaluation. We propose an algorithm that efficiently interpolates between these two possibilities. Our approach elucidates the connection between quantum circuit simulation and partial evaluation of expressions in tensor algebra. 
\end{abstract}


\newcommand{\add}[1]{{\color{blue} #1}}

\maketitle

\section{{Introduction}
\label{sec:introduction}}

An intense interest in quantum computing in recent years lead to the increase of size and capabilities of experimental quantum computers. Promising physical realizations of quantum computing devices were proposed in recent years,~\cite{IntelQ, IBMQ} which bolsters the expectations that a long thought quantum supremacy will be reached.~\cite{harrow2017quantum, boixo2018characterizing, neill2018blueprint}

In the meantime, substantial progress has been made in understanding quantum computation and developing classical simulators of quantum circuits. Efficient simulators were developed for highly parallel computers, such as Sunway Taihulight.~\cite{li2018quantum} At the moment the simulation software is aimed at either one of two tasks. The first one is predicting the probability of measuring a particular binary string as the result of a quantum program, or single amplitude simulation. The second is obtaining the full distribution of quantum circuit outputs, or full state simulation. The first approach was found more economical in terms of memory a classical computer has to use, thus allowing the simulation of few amplitudes of larger quantum circuits on up to 100 qubits.~\cite{Chen2018} On the other side, the second approach may be preferred when the full state information is needed, such as in Shor's algorithm.~\cite{shor1994algorithms}

In this paper, we present a unified approach to quantum circuit simulation. The user can choose the number of probabilities of bitstrings to simulate in a single pass. Our algorithm allows to compromise between the amount of available computational resources and the overall time of the simulation. We build our work on the connection of graphical models and quantum circuits introduced by Markov and Shi~\cite{markov2008simulating} and later developments by Boixo et al.~\cite{Boixo2017} and other authors. Here, we concentrate on an inherently sequential algorithm and defer the discussion of parallelization strategies to a subsequent publication. However, an interested reader is referred to the works \onlinecite{Chen2018, li2018quantum, pednault2017breaking} for efficient parallel simulation algorithms. During the development of this manuscript an interesting work was presented by Pednault et al. in~\onlinecite{pednault2017breaking}. We find that our approach is more straightforward, as it disentangles the problem of multiple amplitude simulation from the parallelization. We defer a more detailed comparison to a later section. An overview of the paper is as follows.

In Sec.~\ref{sec:circuit_simulation}, we review the connection of quantum circuits, tensor diagrams, and statistical graphical models. We then proceed by describing a basic algorithm for circuit simulation based on Refs.~\onlinecite{markov2008simulating, Boixo2017}. In Sec.~\ref{sec:batch_simulation} we formulate the main problem solved in this work, e.g., batch simulation of amplitudes. To solve it, we recall the tree decomposition of graphs and its connection to the problem of ordering of graph nodes. We then propose a new algorithm to transform graph orderings while preserving treewidth (e.g., the quality) of the given ordering. To achieve a proper transformation we use the connection of tree decomposition and chordal graphs, as explained in~\ref{ssec:node_ordering_chordal_graphs} and~\ref{ssec:finding_restricted_order}. Numerical experiments are listed in Sec.~\ref{ssec:numerical_experiments}. Finally, we conclude in Sec.~\ref{sec:conclusion} with final remarks and outline possible future research. Due to the growing body of literature on quantum circuit simulation, we included an additional Sec.~\ref{sec:related_work} in the Appendix for comparison of our approach to other competing techniques.

\section{Related work\label{sec:related_work}}
Here we will put our method in the scope of existing research. The reader interested in implementation is suggested to move to Section \ref{sec:circuit_simulation} and to revisit this section later.

The problem of efficient tensor contraction was approached multiple times in the field of many-body physics and quantum computing. Some older works are based on the sequential application of sparse matrices to the state vector, such as in~\onlinecite{de2007massively}. The authors issued a follow-up paper recently~\cite{de2019massively}. Their simulator can evaluate both full sets and subsets of amplitude tensor. This direct simulation procedure,
however, requires a lot of non-trivial techniques to make it efficient, especially if parallel operation is considered. Another problem
is that it is hard to analyze the effectiveness of the algorithm compared to theoretical bounds on the numerical cost~\cite{aaronson2016complexity}. The latter fact has lead to the previously believed margin of 50 qubits for "quantum supremacy". 

The seminal work of Markov and Shi~\cite{markov2008simulating} introduced tensor networks for quantum algorithm simulations and showed that treewidth is a natural measure of simulation hardness. The graph-based notation became standard in tensor network literature a decade ago.~\cite{bridgeman2017hand} Following Markov and Shi, several groups developed highly efficient algorithms for quantum circuit simulation based on this representation, see~\onlinecite{pfeifer2014faster, chen201864, pednault2017breaking, li2018quantum} for more details. The previous margin of 50 qubits was lifted, as is demonstrated by multiple authors.~\cite{chen201864, pednault2017breaking, li2018quantum} Usually, these simulators are capable of evaluating full state vectors as well as some subsets of the amplitudes. A similar program was created for contraction of tensors emerging in the many-body physics community~\cite{pfeifer2014faster}. The drawback of the approaches based on tensor diagrams is the hardness of the development of efficient codes and the theoretical performance analysis, especially if parallelization is involved. To see why, let us note that classical tensor networks were developed to represent pairwise contractions. Quantum circuits
often involve multiple diagonal gates, which allows for significant computational savings. The treewidth of
classical diagram's graphs is higher than optimal (see Appendix in~\onlinecite{Boixo2017}). Traditional network notation can be understood as a hypergraph to eliminate this drawback, as was done in~\onlinecite{pednault2017breaking}. However, the theory of hypergraphs is less known to the general scientific community. 

Recently Boixo et al.~\cite{Boixo2017} proposed to consider line graphs of the classical tensor networks, which has multiple benefits. First, it establishes
the connection of quantum circuits with probabilistic graphical models, allowing for knowledge transfer between the fields. Second, these graphical
models avoid the overhead of traditional diagrams for diagonal tensors. Moreover, the treewidth is a universal measure of complexity for these models, and links the complexity of quantum states to the well-studied problems in graph theory, a topic we hope to explore in future works.
Additionally, simple parallelization of the simulator is possible, as demonstrated in the work of Chen et al.~\cite{Chen2018} The only disadvantage of the
line graph approach was that it is limited usability to simulate subtensors of amplitudes, which we are going to fulfill in this article.

\section{Quantum circuit simulation algorithm \label{sec:circuit_simulation}}
In this section we describe a procedure for efficient quantum circuit evaluation.
We first set up the notation, and then review the current state of the art method for numerical simulation of quantum circuits.

\subsection{Tensor networks and graphical models
\label{ssec:tensor_nets}}
A quantum program describes an evolution of the initial state $\ket{0}$ of a system of $n$ qubits. Any evolution of a physical system corresponds to a unitary operator. Thus, the result of a quantum circuit is a state $|\psi \rangle$, which is a linear transformation of the input state: $|\psi \rangle = \mathcal{U}|0\rangle$. Usually, the transformation $U$ is performed in several steps corresponding to clock cycles of a quantum computer. Let us introduce the following notation:
\begin{equation}
\begin{aligned}
&\mathcal{U} \ket{0} = \mathcal{U}^{d} \dots \mathcal{U}^{2} \mathcal{U}^{1} \ket{0} \\
& \ket{s^{t+1}} = \mathcal{U}^{t} \ket{s^{t}}, \quad \ket{s^{0}} = \ket{0}
\end{aligned}
\end{equation}
Here $U_{t}$ are unitary matrices acting at the $t$-th clock cycle and $\ket{s^{t}}$ is the state vector. In the simplest case the initial state is taken to be a product of single qubit states $\ket{0} = \ket{0_{0}} \otimes \dots \otimes \ket{0_{n}}$. A naive simulation algorithm would take the initial vector $\ket{0}$ and apply matrices $\mathcal{U}^{t}$ to it. This procedure lays behind full state circuit simulation. To calculate an amplitude of a bit string $x$, one would evaluate a dot product $\braket{x|s^{d}}$:
\begin{equation}
\begin{aligned}
\sigma(x) = \braket{x | s^{d}} = \sum_{i=1}^{n} \braket{x_i | s^{d}}
\end{aligned}
\label{eq:dot_product_amplitude}
\end{equation}
The probability of $x$ is then the modulus squared of the amplitude. Note, however, that it is hard to perform full state simulation efficiently. 
A naive algorithm would need to operate on vectors of size $2^n$. Also, the matrices $\mathcal{U}^{t}$ are highly sparse, at least if they represent transformations achievable with single and two-qubit gates in modern experimental hardware.
Here and later in the paper, we chose to work with the following universal set of one and two-qubit gates: $\{X^{1/2}, Y^{1/2}, cZ, T, H\}$; the same reasoning, however, applies to any quantum gates.

\begin{figure*}
\includegraphics[width=0.7\textwidth]{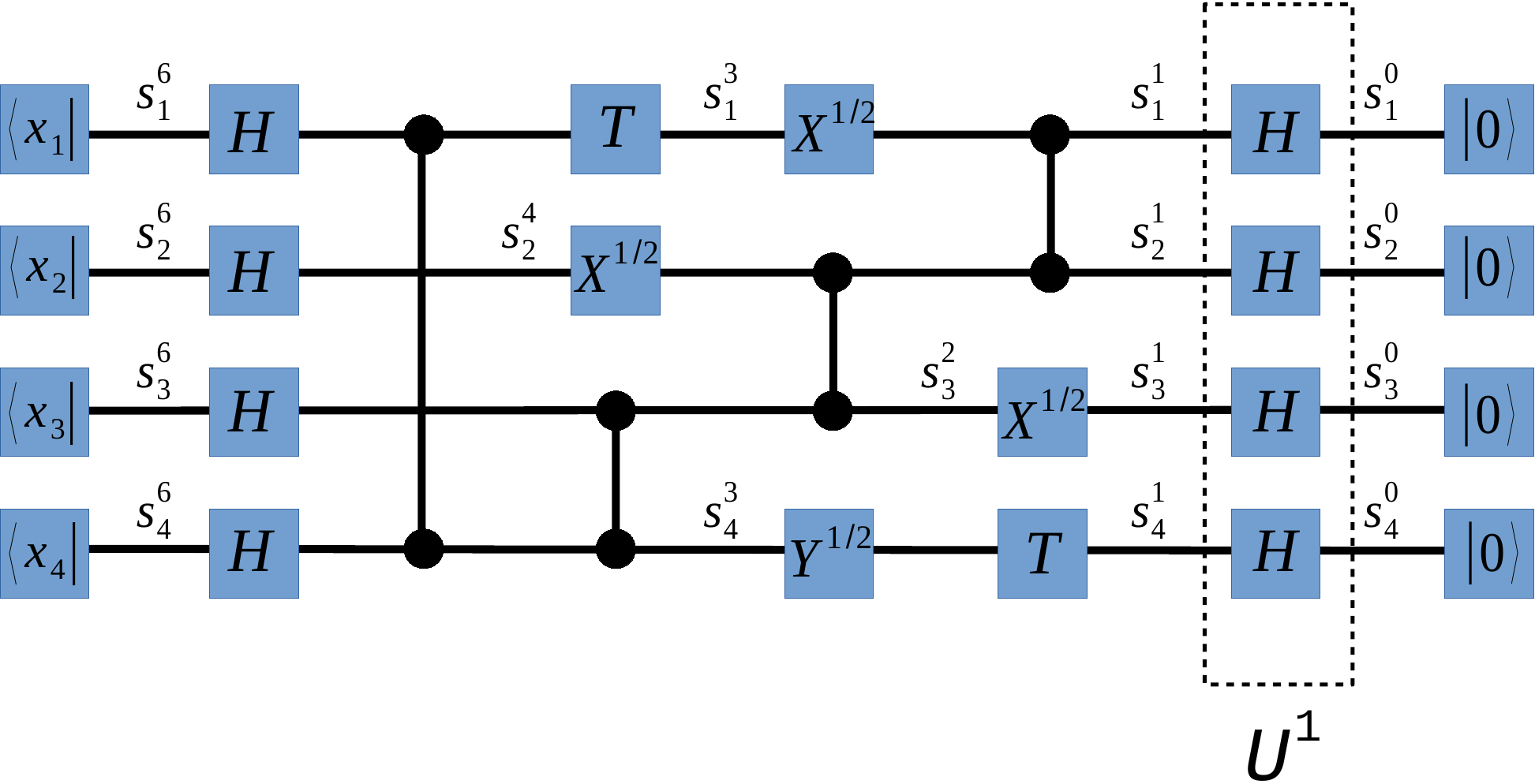}
\caption{Example of quantum circuit drawn as a tensor network. The state of $i$-th qubit at $t$-th clock cycle is denoted by $\{s\}_{i}^{t}$ (only unique states are shown, e.g. $s_{1}^{2} = s_{1}^{1}$ is omitted) \label{fig:circuit}}
\end{figure*}

An alternative to full state simulation would be the evaluation of one or several amplitudes from Eq.~\ref{eq:dot_product_amplitude} without explicitly forming $\ket{s^d}$. The latter approach provides several benefits. First of all, we can avoid storing the high dimensional state vector $\ket{s^d}$ in computer memory. Second, it may be easier to use the internal structure of the operators $\mathcal{U}^{t}$ to perform calculations efficiently. Let us introduce a set of variables to denote the state at different cycles of the circuit.
\begin{equation}
\begin{split}
& \{ s \}^{t}_{i}, \quad s \in [0, 1], \quad i \in [1, n], \quad t \in [0, d]
\end{split}
\end{equation}
We slightly abuse notation here, as $\ket{s_{i}^{t}}$ denotes a state of the $i$-th qubit at $t$-th cycle, and $s_{i}^{t}$ is a binary
variable indexing this state. Same notation is used for the initial and final states, e.g. $\ket{s_{i}^{0}} = \ket{0_{i}}$ and $\ket{s_{i}^{d}} = \ket{x_{i}}$. Consider a circuit shown in Fig.~\ref{fig:circuit}.

\onecolumngrid
\twocolumngrid

We start with a product state $\ket{0}$ on the right. As the program proceeds the states of individual qubits are changed by gates application. Note that the gates $T$ and $cZ$ do not change the basis of the single-qubit subspaces they act on (they only multiply basis vectors by constants), and hence $\ket{s_{i}^{t}} = \ket{s^{t+1}_{i}}$ for those qubits. In contrast, non-diagonal gates $\{X^{1/2}, Y^{1/2}, H\}$ mix basis vectors of the appropriate qubit subspaces, and new variables $\ket{s^{t+1}_{i}}$ have to be introduced for the resulting bases. On Fig.~\ref{fig:circuit} only unique variables are shown. The expression for the single amplitude in Eq.~\ref{eq:dot_product_amplitude} can be rewritten as
\begin{equation}
\begin{aligned}
& \sigma(x) = \braket{x | \mathcal{U} | 0} = \\
\sum_{\{s_{i}^{t}\}} \braket{x_i | \mathcal{G}^{d}_{i} | s^{d-1}_{i} } & \ldots \braket{s^{t+1}_i s^{t+1}_j | \mathcal{G}^{t}_{ij} | s^{t}_{i} s^{t}_{j}} \ldots \braket{s^{1}_{i}| \mathcal{G}^{1}_{i} | 0_{i}}\\ 
\mathcal{G}^{t}_{i} \in \{X^{1/2}, Y^{1/2},& T, H\}, \qquad \mathcal{G}^{t}_{ij} = cZ
\label{eq:amplitude_extended}
\end{aligned}
\end{equation}
The Eq.~\ref{eq:amplitude_extended} can be interpreted as a discrete Feynman path integral. On the other hand, one can easily see that the evaluation of the amplitude $\sigma(x)$ in Eq.~\ref{eq:amplitude_extended} is equal to the contraction of a tensor network shown in Fig.~\ref{fig:circuit} (for the introduction to the graphical notation used for tensor networks please refer to \onlinecite{cichocki2016tensor}). It is well known, however, that the numerical cost of tensor contractions dramatically depends on the order of operations. Following Markov and Shi~\cite{markov2008simulating} let us introduce another type of graphical models to denote quantum circuits, which better suits for the estimation of numerical costs.

\begin{figure}
\includegraphics[width=0.45\textwidth]{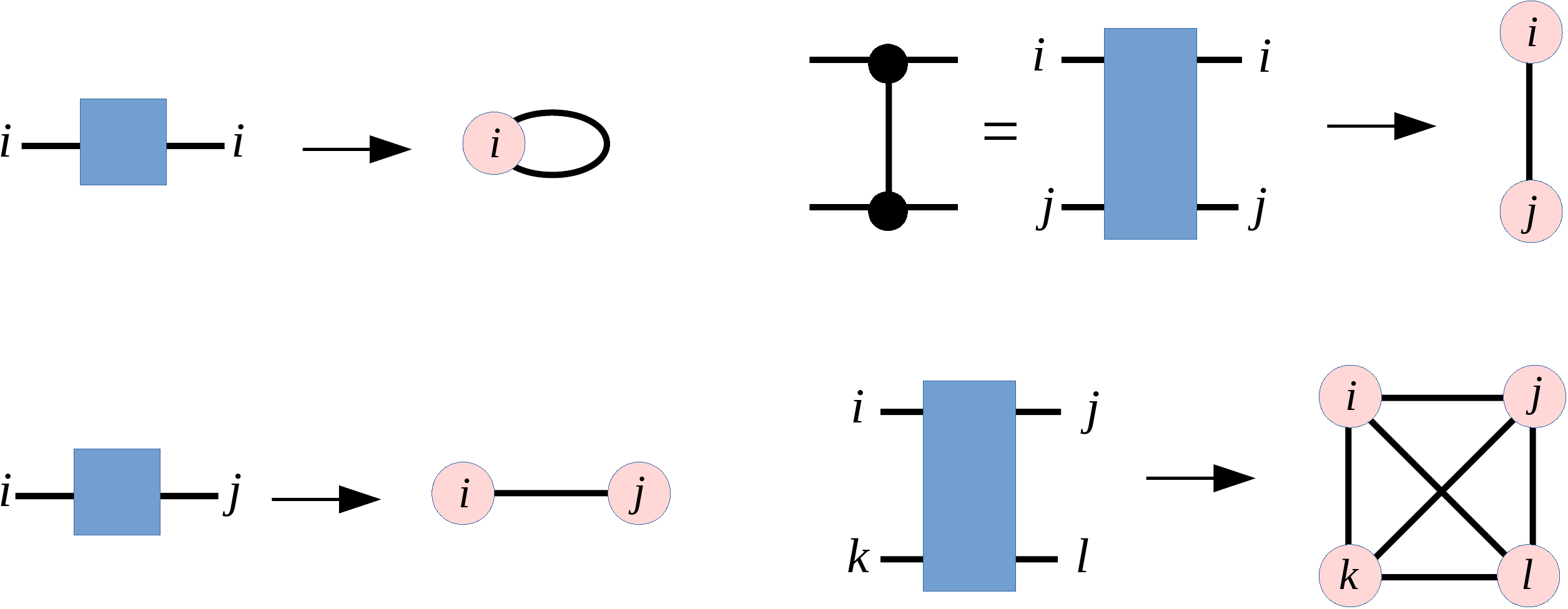}
\caption{The mapping between two graph-based notations of tensor networks. \label{fig:gadgets}}
\end{figure}

In traditional notation, a tensor network is represented by a graph with nodes standing for tensors and edges denoting their indices. In the new notation, we use nodes to denote unique indices, and tensors are denoted by cliques (fully connected subgraphs). Note that tensors, which are diagonal along some of the axes and hence can be indexed with fewer variables, are depicted by cliques of size lower than the dimension of the corresponding tensor. For a special case of vectors or diagonal matrices, self-loop edges are used. Fig.~\ref{fig:gadgets} lists the notation for the gates used in this work.

\begin{figure}
\includegraphics[width=0.45\textwidth]{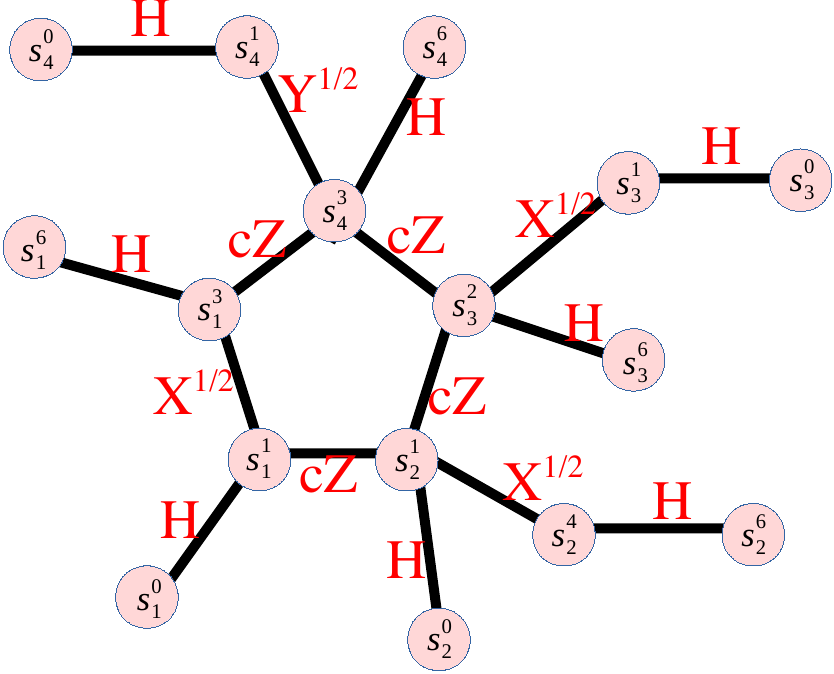}
\caption{Alternative representation (graphical model) of the circuit in Fig.~\ref{fig:circuit}. Gate tensors are shown in red, self-loops are omitted.} \label{fig:graph_model}
\end{figure}

A graphical model, which is equivalent to the circuit in Fig.~\ref{fig:circuit}, is shown in Fig.~\ref{fig:graph_model} (self loops are omitted for simplicity). As was pointed out by Boixo et al.\cite{Boixo2017} this representation of tensor contractions is traditional in Bayesian network literature. Notice that provided a quantum circuit in a traditional form, one can easily build its probabilistic model representation. To do that, one has to replace all edges carrying non-equivalent single-qubit states with nodes, and all gates with cliques. The diagonal structure of $cZ$ gate tensors leads to significant simplification of the resulting graphs.

\subsection{ simulation of quantum circuits}
Having set up the notation, let us proceed with a description of a basic procedure for the evaluation of tensor networks. This algorithm was developed in the context of probabilistic models under the names of bucket elimination\cite{detcher2013bucket} or the variable elimination algorithms\cite{marsland2011machine}. 

\begin{figure}
\includegraphics[width=0.45\textwidth]{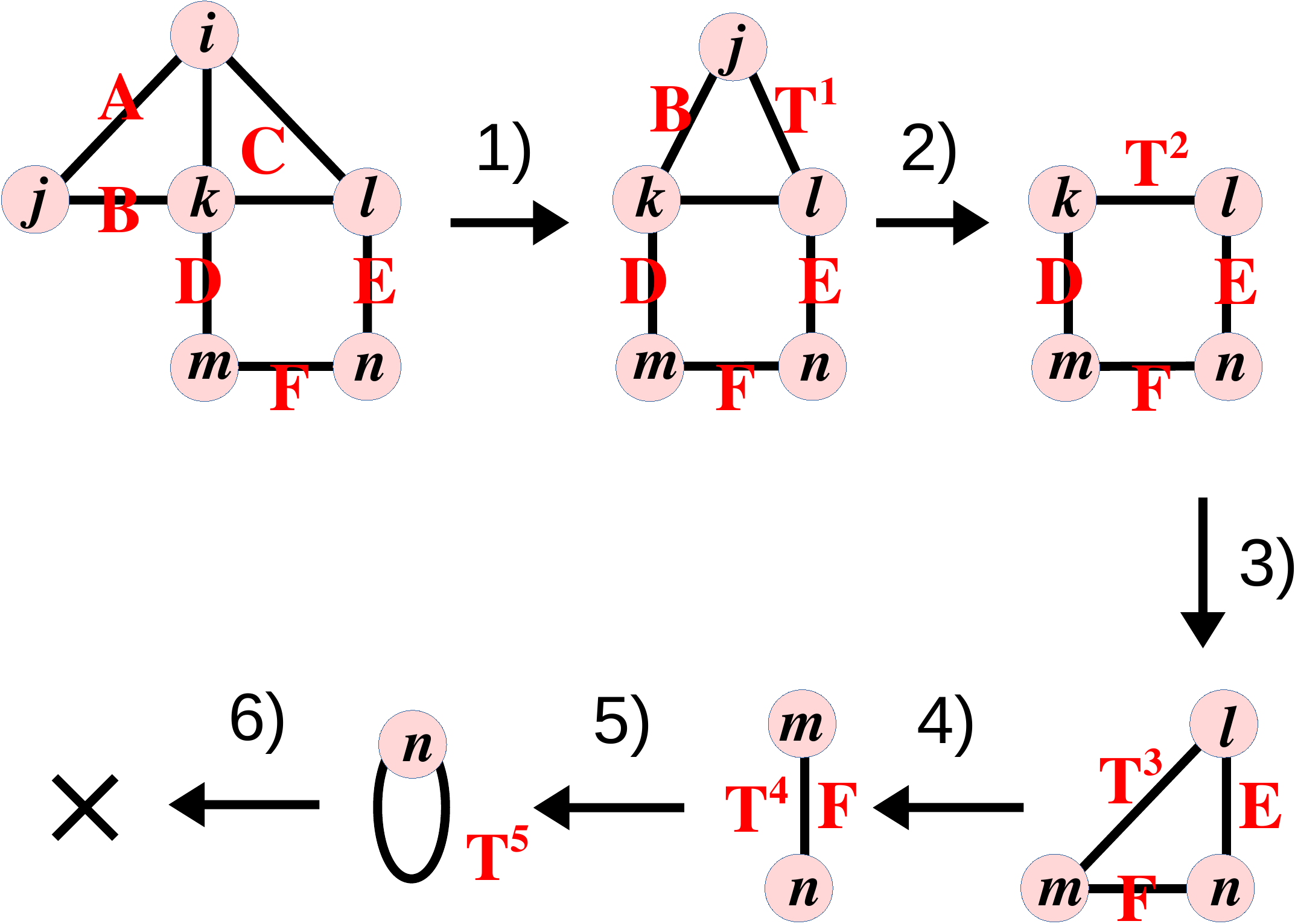}
\caption{Contraction of a tensor network from Eq.~\ref{eq:simple_graph_model} in graphical form. The sequence of contractions $\pi$ is the same as in Eq.~\ref{eq:sequence}. \label{fig:simple_graph_model}
Labels of tensors are shown in red.}
\end{figure}

As an example, let us consider the contraction of a simple tensor network:
\begin{equation}
\sum_{ijklmn} A_{ij} B_{jk} C_{ikl} D_{km} E_{ln} F_{mn} = \sigma
\label{eq:simple_graph_model}
\end{equation}
The graphical model of this network is shown in Fig.~\ref{fig:simple_graph_model}. We choose the order
of indices as $\pi = \binom{i ~ j ~ k ~ l ~ m ~ n}{1 ~ 2 ~ 3 ~ 4 ~ 5 ~ 6}$, e.g. $i$ is first, $j$ is second etc. In bucket elimination procedure the indices are contracted one at a time in order fixed by $\pi$, until no indices is left. The sequence of the expressions evaluated in the algorithm is listed below. Assuming the dimensions of all indices is $L$, we also list numerical costs of the operations.
\begin{equation}
\begin{split}
& 1) ~\sum_{i} A_{ij} C_{ikl} = T^{1}_{jkl} \quad \mathcal{O}(L^4) \\ 
& 2) ~\sum_{j} B_{jk} T^{1}_{jkl} = T^{2}_{kl} \quad \mathcal{O}(L^3) \\
& 3) ~\sum_{k} D_{km} T^{2}_{kl} = T^{3}_{ml} \quad \mathcal{O}(L^3) \\
& 4) ~\sum_{l} E_{ln} T^{3}_{ml} = T^{4}_{nm} \quad \mathcal{O}(L^3) \\
& 5) ~\sum_{m} T^{4}_{nm} = T^{5}_{n} \quad \mathcal{O}(L^{2}) \\
& 6) ~\sum_{n} T^{5}_{n} = \sigma \quad \mathcal{O}(L)
\end{split}
\label{eq:sequence}
\end{equation}
The sequence of transformations of the graphical model corresponding to Eq.~\ref{eq:sequence} is shown in Fig.~\ref{fig:simple_graph_model} 2) - 6).

At each step, the contracted variable is removed from the graph, and all its neighbors form a clique. This clique corresponds to the next intermediate in the sequence. Note that the order of the cliques formed at each step corresponds to the exponent of the scaling of numerical cost.

\begin{figure}[hbt]
\includegraphics[width=0.45\textwidth]{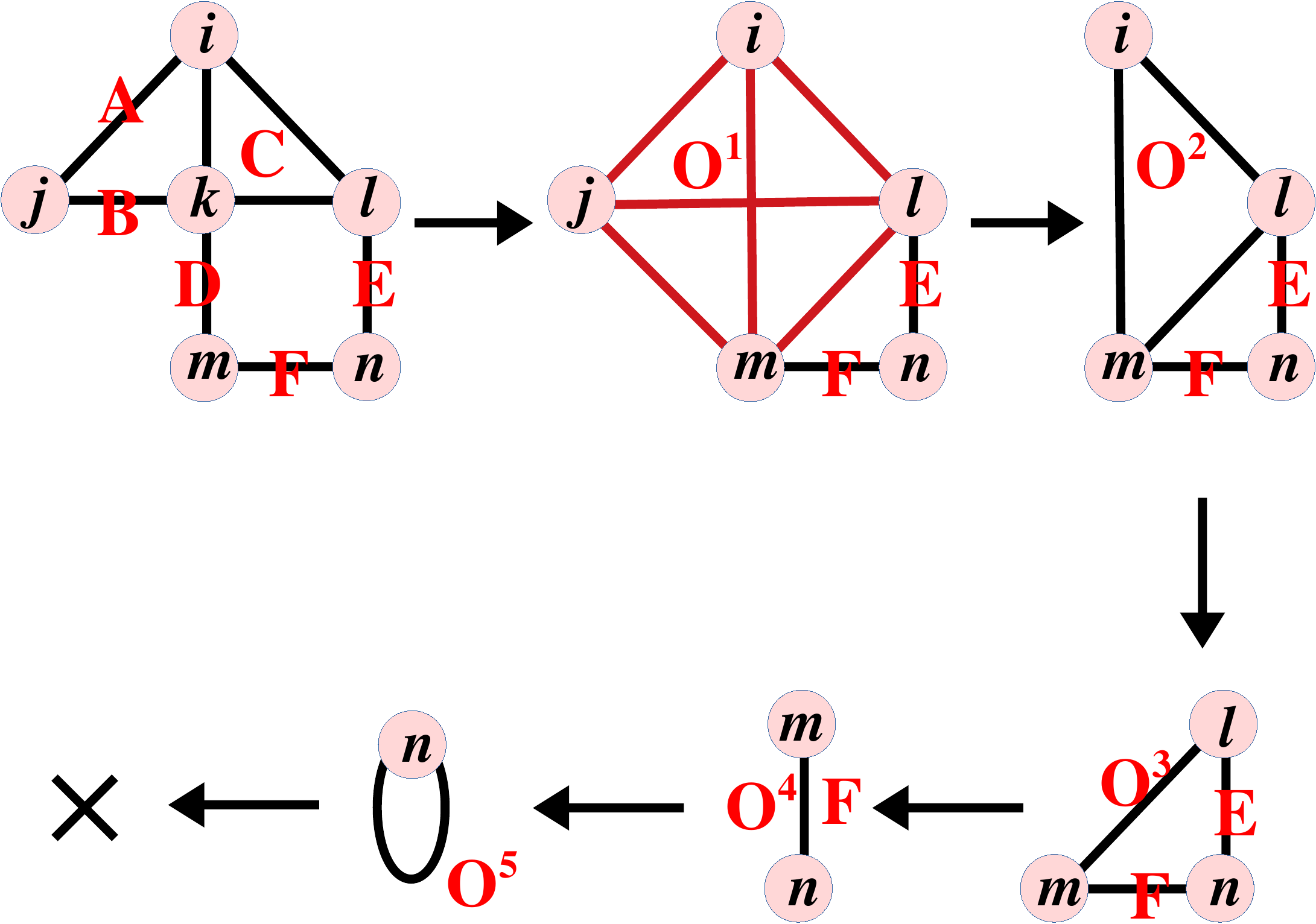}
\caption{Alternative contraction of a tensor network in Eq.~\ref{eq:simple_graph_model}. The maximal clique of size 4 is highlighted in red. This sequence of contractions is not optimal.\label{fig:simple_graph_model2}}

\end{figure}

The computational cost of the tensor network contraction is highly dependent on the order of operations. To illustrate this let us consider an alternative order $\tilde{\pi} = \binom{k~j~i~l~m~n}{1~2~3~4~5~6}$ for evaluating the Eq.~\ref{eq:simple_graph_model}. The corresponding sequence of graphical models is shown in Fig.~\ref{fig:simple_graph_model2}. Note that the size of the maximal clique corresponding to order $\tilde{\pi}$ is four, which translates to the intermediate of order four and the overall scaling $\mathcal{O}(L^5)$ of the numerical effort. 

Finding the elimination order of a graph is equivalent to the calculation of its tree decomposition; the size of the maximum clique of an order $\pi$ is $\mathrm{treewidth} + 1$. Tree decomposition is NP-hard for general graphs~\cite{bodlaender1994tourist}, and a similar hardness result is known for the optimal tensor contraction problem~\cite{chi1997optimizing}. However, several exact and approximate algorithms for tree decomposition were developed in graph theory literature; for references, please see \onlinecite{gogate2004complete, bodlaender2006exact, kloks1994treewidth, bodlaender1994tourist, kloks1993computing}. For our simulations, we used an exact algorithm of V. Gogate\cite{gogate2004complete}. Having reviewed the procedure for calculation of a single amplitude, let us consider the case of multiple amplitudes, which is the main topic of this article.

\section{Batch circuit simulation \label{sec:batch_simulation}}
\subsection{Simulation of multiple amplitudes \label{ssec:simulation_of_multiple_amplitudes}}

\begin{figure}
\includegraphics[width=0.45\textwidth]{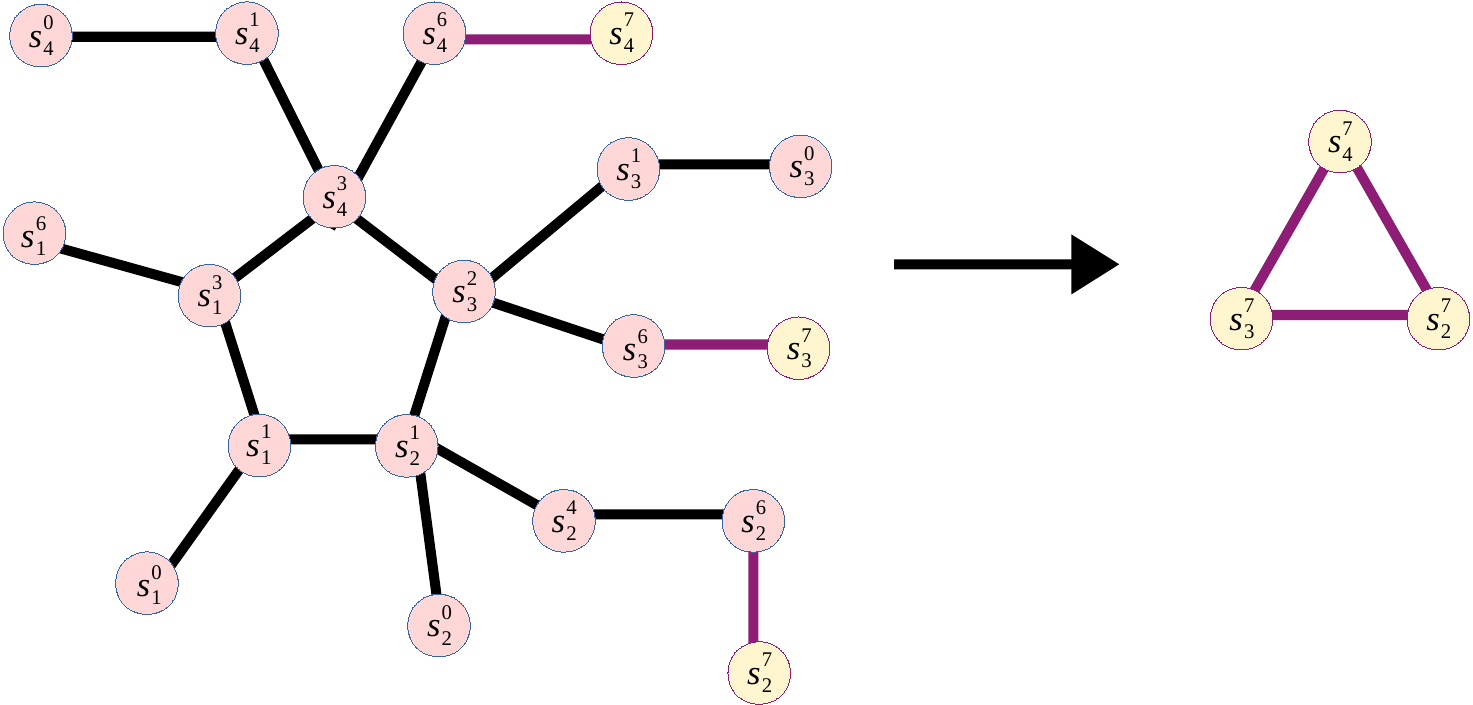}
\caption{Evaluation of amplitude subsets}
Left - extended amplitude expression to evaluate all amplitudes of qubits 2, 3, 4;
Right - resulting amplitude tensor.
\label{fig:graph_model_multiple}
\end{figure}

The procedure we used to calculate single amplitude can be easily extended to calculate any subtensor of the full amplitude tensor.
Suppose we are interested in amplitudes of two bitstrings differing only in the value of the first qubit, e. g. $x^{0} = (0, s^{d}_{2}, \dots, s^{d}_{n})$ and $x^{1} = (1, s^{d}_{2}, \dots, s^{d}_{n})$. Let us note that the expressions for the amplitudes of $\sigma^{0}$ and $\sigma^{1}$ differ only by the value of the state vector of the first qubit, which is $\ket{0}$ and $\ket{1}$ respectively.
One could merge both expressions and introduce an additional variable $s^{d+1}_{1}$ to index the result $\sigma(s^{d+1}_{1})$ (which is a vector of size two). The same procedure can be implemented for any combination of output qubits; thus, any subtensor of the full amplitude tensor can be encoded.
A graphical representation of the extended amplitude expression is shown in Fig.~\ref{fig:graph_model_multiple}. We have to mention that the same procedure can be used not only to evaluate the probabilities of multiple output states, but also the evolution of multiple input states. This approach can be used to simulate the dynamics of mixed states, although we will not elaborate on this in the current article.

In order to evaluate multiple amplitudes, the resulting extended expressions have to be contracted only partially (the indices of the amplitude subtensors should not be summed over). Partial contraction can be achieved by stopping the bucket elimination algorithm when all necessary indices are eliminated and (possibly) merging the final set of tensors. Notice that the result of the evaluation of all amplitudes for $c$ qubits will result in a tensor with $2^c$ elements, which will be mirrored by a clique (a fully connected subgraph) with $c$ nodes in our graphical notation (Fig.~\ref{fig:graph_model_multiple}, right). 

After selecting a subset of nodes to leave in the result, one still faces a problem of choosing an optimal order of variable elimination to implement partial contractions. Let us turn to the discussion of a possible solution.

\subsection{Node ordering and chordal graphs
\label{ssec:node_ordering_chordal_graphs}}
To properly introduce the procedure of finding elimination orders for partial contractions, let us first highlight the connection of
elimination orders and chordal graphs. Chordal graphs (also called triangulated graphs) are the ones that do not have cycles of length higher than 3. Many problems, which
are hard on general graphs can be solved on chordal graphs in polynomial time (for example, the Maximum Clique problem\cite{gavril1972algorithms}). An extensive introduction to the properties of chordal graphs and related algorithms can be found in \cite{blair1993introduction}.

We will employ chordal graphs because of their relation to node orderings. Consider the bucket elimination procedure described before, but without node removal. Specifically, given a graph $G$ and a node order $\pi$,
one would loop over the nodes according to $\pi$ and for each node connect all of its neighbors who have higher order in $\pi$. It can be shown\cite{blair1993introduction} that this procedure will always produce a chordal graph. Indeed, if the initial graph had any cycle with four or more nodes, connecting the neighbors of any node in the cycle will introduce a chord, thus breaking a cycle into smaller, three node cycles. The resulting chordal graph is also called a fill-in graph in this context. 

\begin{figure}
\includegraphics[width=0.45\textwidth]{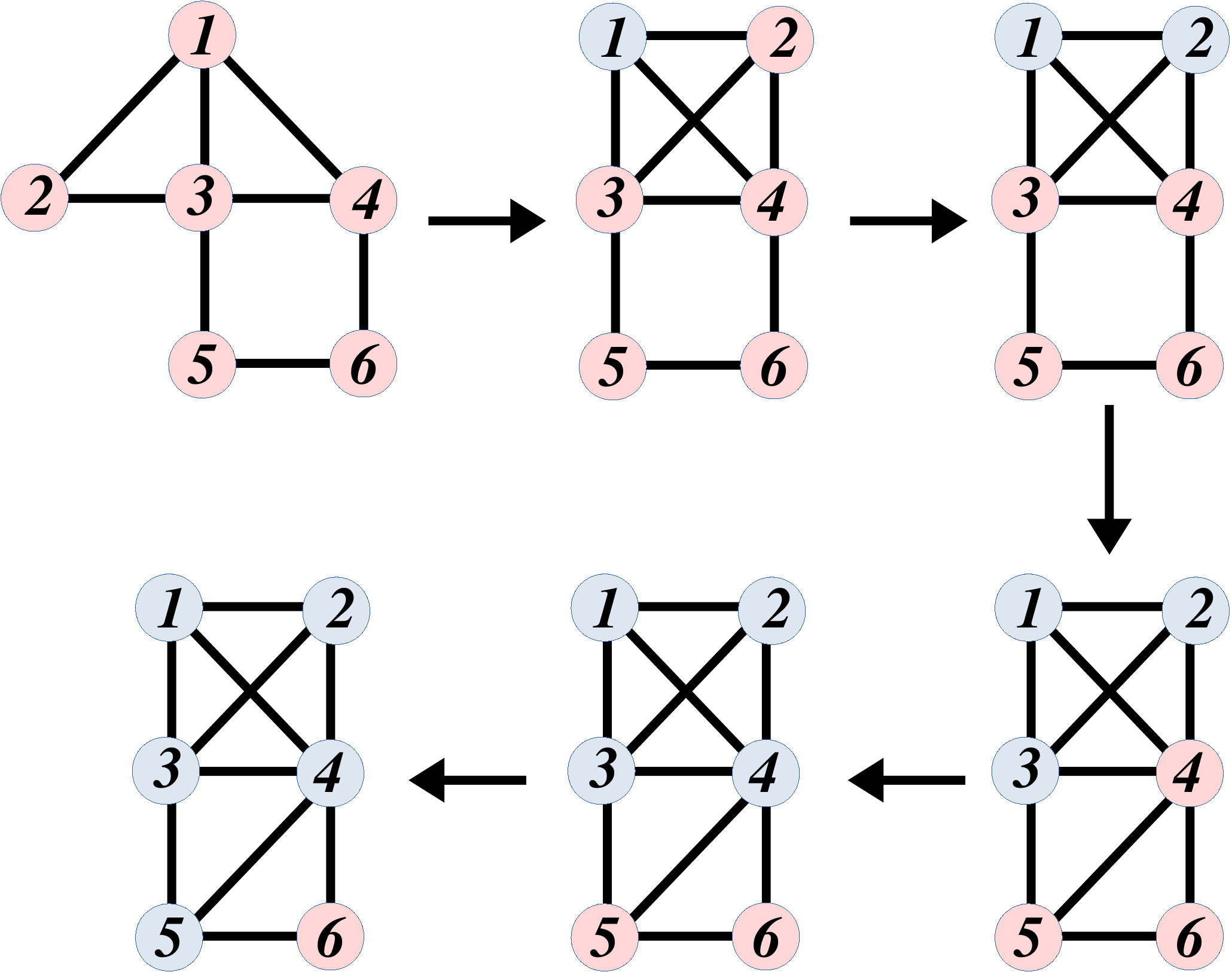}
\caption{Building a chordal graph from the elimination order}
The graph corresponds to the tensor network in Eq.~\ref{eq:simple_graph_model}.
The nodes are labeled according to their order.
\label{fig:chordal_graph}
\end{figure}

A formal algorithm for building a fill-in graph $H$ given an initial graph $G$ and an elimination order $\pi$ is listed in Alg.~\ref{code:build_chordal_graph}. A corresponding graphical representation of the algorithm is provided in Fig.~\ref{fig:chordal_graph}. An important remark has to be made here. Any elimination order $\pi$ will produce a chordal graph, but this does not imply that there is a one-to-one correspondence. Multiple orders can result in the same fill-in graph \cite{tarjan1984simple}; we will employ this fact in the next section.

\begin{algorithm}[H]
\caption{Building chordal graph from the elimination order}
\label{code:build_chordal_graph}
\begin{algorithmic}[1]
\Require $G = (V, E), \pi: V \rightarrow \mathcal{N}, ~~ \pi = \{(v_{i}, i)\}_{i=1}^{|V|}$
\Ensure $H = (V, \tilde{E}), ~~ \mathrm{H ~ is ~ chordal}$
\Statex
\Function{Build\_chordal\_graph}{$G, \pi$}
\State $\tilde{E} \gets E$
\For{$i \in [1, \ldots, |V|]$}
\State $v \gets \pi^{-1}(i)$
\State $U = \emptyset$
\For{$w ~ \mathrm{in ~ neighbors}(v)$}
\If{$\pi(w) > i$}
\State $U \gets U \cup w$
\EndIf
\EndFor
\For{$x, y ~ \mathrm{in ~ pairs}(U)$}
\State $\tilde{E} \gets \tilde{E} \cup (x, y)$
\EndFor
\EndFor
\EndFunction
\end{algorithmic}
\end{algorithm}

The size of the maximum clique in the fill-in graph equals \emph{treewidth} by construction \cite{bodlaender2006exact}.
The problem of searching the best elimination order for a graph $G$ thus can be formulated in terms of the search of an
optimal fill-in graph. Formally, given a graph $G = (V, E)$ the task of finding an elimination order $\pi$ with minimal treewidth is
equivalent to finding a chordal graph $H = (V, \tilde{E}), ~~ \tilde{E} \in E$, such that the size of its maximum clique is minimized.

Provided a chordal graph $H$ is found, any of its elimination orders that does not introduce additional edges, and hence does not change the graph $H$, will have the same treewidth. Chordal graphs thus provide means of building equivalent (in terms of treewidth) elimination orders. In contrast with arbitrary graphs, finding elimination orders of chordal graphs can be done in linear time \cite{tarjan1984simple}. We now turn to the description of the procedure of building of equivalent elimination orders of chordal graphs.

\subsection{Finding restricted elimination orders
\label{ssec:finding_restricted_order}}
Let us now find an optimal elimination order for multiple amplitude evaluation, as described in Sec.~\ref{ssec:simulation_of_multiple_amplitudes}. In essence, we want to find an order with minimal treewidth, such that some set of nodes will be at the end of this order. Putting it formally, for a graph $G = (V, E)$ and a subset of nodes $C \in V$ we want to find an order $\pi$ with minimal treewidth, such that for any nodes $v \in C$ and $w \in V \setminus C$ the order of $v$ is higher than the order of $w$: $\pi(v) > \pi(w)$. 

Our idea is to calculate an optimal unrestricted elimination order $\tilde{\pi}$ (not necessary having $C$ at the end), and then to use the connection between the elimination orders and chordal graphs to transform it to the desired order $\pi$. Essentially, we employ the result of Bodlaender~\cite{bodlaender2006exact}, 
to devise a procedure for building $\pi$. Our approach is outlined below:

\begin{enumerate}
\item Check if $C$ induces a clique in $G$. If $G[C]$ is not a clique, turn $G[C]$ into a fully connected subgraph. This step ensures that the condition stated in (Ref.~\onlinecite{bodlaender2006exact}, Lemma 10) is satisfied: if $C$ is a clique, then there always exists an elimination order with $C$ at the end. A graph $\tilde{G}$ is produced as a result of (possibly) turning $C$ into a clique.
\item Find an elimination order $\tilde{\pi}$ of $\tilde{G}$ using an exact (NP-hard) or a heuristic algorithm. We use the branch and bound algorithm of Gogate\cite{gogate2004complete} with the time limit of 60 seconds (to obtain an exact solution the algorithm has to be given a very long time).
\item Build a chordal graph $H$ using Alg.~\ref{code:build_chordal_graph}.
\item Provided with a set $C$ and a chordal graph $H$, construct a new order $\pi$ with the help of the
Restricted Maximum Cardinality Search (MCS) algorithm. The order $\pi$ has same treewidth as the order $\tilde{\pi}$
and nodes in $C$ are placed at the end in $\pi$.
\end{enumerate}

Essentially, in our approach, we transform an arbitrary solution to the Tree decomposition problem to the one that satisfies
our restrictions (places all nodes in $C$ to the end) and has the same quality (same treewidth). The last ingredient to complete the procedure is the Restricted Cardinality Search algorithm. We modified the original algorithm from Ref.~\onlinecite{tarjan1984simple}. The resulting pseudocode is provided in
Alg.~\ref{code:mcs_algo}.

\begin{algorithm}[H]
\caption{Computing an elimination order with a set of nodes $C$ at the end}\label{code:mcs_algo}
\begin{algorithmic}[1]
\Require $H = (V, E), ~ H ~\mathrm{is ~ chordal}, ~~ C \in V, ~ C ~\mathrm{is ~ clique}$
\Ensure $\pi: V \rightarrow \mathcal{N}, ~~ \pi = \{(v_{i}, i)\}_{i=1}^{|V|}$
\Statex
\Function{Restricted-MCS}{$H, C$}
\For{$v \in V$}
\State cardinality($v$) $\gets 0$
\EndFor
\For{$i \in [|V|, |V| - 1, \dots, 1]$}
\If{$C \neq \varnothing$}
\State pick $v \in C$, $C \gets C \setminus \{v\}$
\Else
\State pick $v \in V$ with maximum cardinality
\State $V \gets V \setminus \{v\}$
\EndIf

\State $\pi \gets \pi \cup \left(v, i \right)$

\For{$w \in$ neighbors($v$), $w \notin \pi$}
\State cardinality($w$) $\gets$ cardinality($w$) + 1
\EndFor

\EndFor
\EndFunction
\end{algorithmic}
\end{algorithm}

In the Alg.~\ref{code:mcs_algo} each node $v$ of the graph $H$ is assigned a counter "cardinality", which holds the number of
labeled neighbors of $v$. At each step, we label the next node with maximal cardinality, breaking ties arbitrarily.
The order is built in a reversed form. In the beginning, nodes in $C$ are labeled (to be last), and then the rule stated before is applied.
Note that if at step $i$ a node $v$ is selected, then in the next steps all neighbors of $v$, which belong to the maximal clique $K, ~ v \in K$ will be labeled. Overall, the procedure in Alg.~\ref{code:mcs_algo} is polynomial in the number of nodes $|V|$.

Let us now demonstrate the benefits of the proposed approach with numerical examples.

\subsection{Numerical examples
\label{ssec:numerical_experiments}}
The methods developed in previous sections were used to implement a quantum circuit simulator. As numerical examples
we use the simulation of random quantum circuits from the work of Boixo et al. The qubits are arranged in a square grid
of size $k \times k$ and a set of gates $\{X^{1/2}, Y^{1/2}, cZ, T, H\}$ is applied in a predefined pattern. This circuit choice
can be implemented in superconducting quantum processors.~\cite{chen2014qubit} The reader is referred to~\onlinecite{Boixo2017} to learn
more details about the motivation of these random circuits. The dataset with random circuits of Boixo et al., which we used in this work, is available online \cite{boixocircuits}. Here we are interested only in the numerical cost and the memory usage to evaluate amplitudes.

\begin{figure}[hbt]
\includegraphics[width=0.50\textwidth]{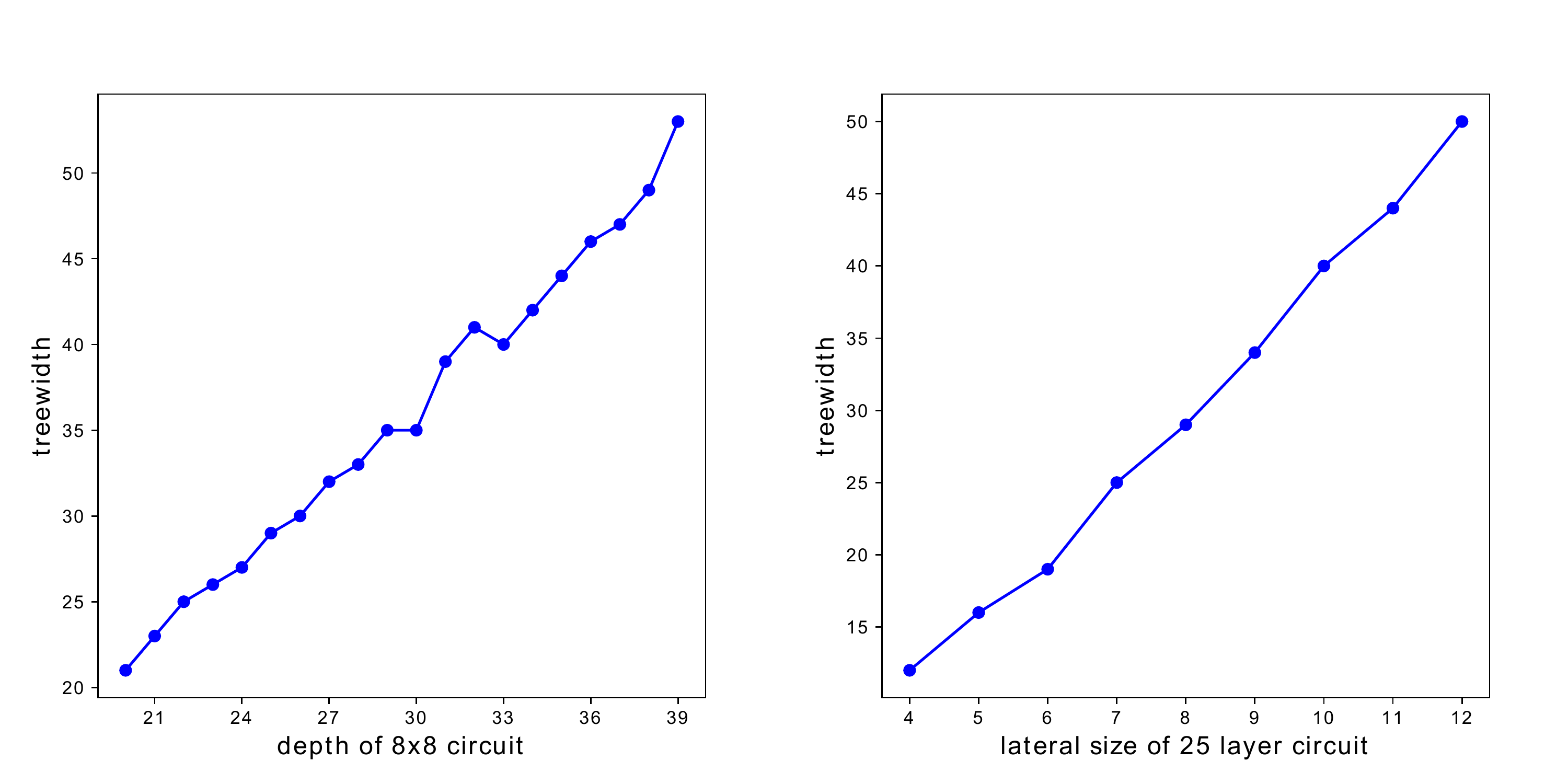}
\caption{Treewidth dependence on the size of a random quantum circuit}
Left - dependence of treewidth on the depth of a random circuit, Right - dependence of treewidth on the number of qubits. 
\label{fig:tw_dependence}
\end{figure}

In Fig.~\ref{fig:tw_dependence}, the dependence of treewidth $\nu$ on the size and depth of the random circuits is shown. Let us recall that the number of floating-point operations scales as $O(2^{\nu+1})$ and the required memory as $O(2^{\nu})$. The complexity of simulation grows exponentially
with the volume of random quantum programs, which was the original motivation for considering them as a test bench for demonstrations of "quantum supremacy".\cite{boixo2018characterizing}

\begin{figure*}[hbt]
\includegraphics[width=0.95\textwidth]{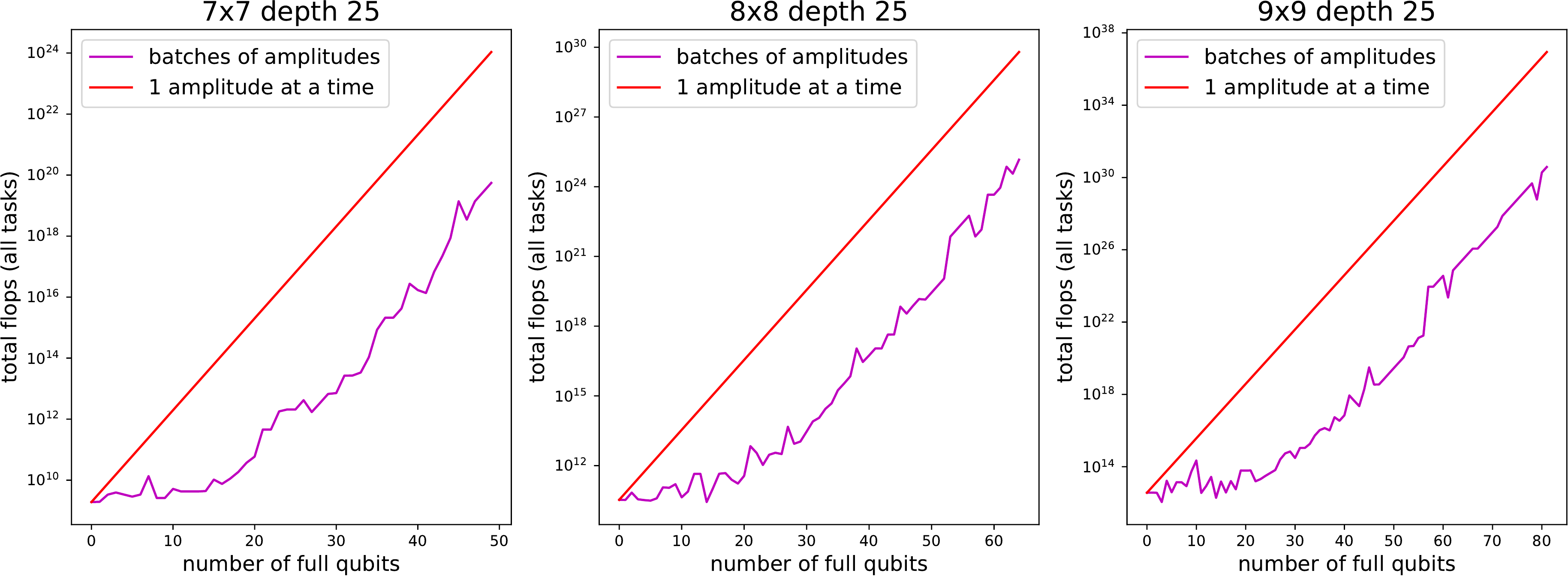}
\caption{Total flop requirements for the simulation of a typical random circuit of varying size}
Shown is the predicted number of floating point operations for the simulation of the full subset of amplitudes of $|C|$ qubits. In case of one amplitude at a time simulation a combined cost all tasks is drawn. 
\label{fig:flops_vs_batch_size}
\end{figure*}

\begin{figure*}[hbt]
\includegraphics[width=0.95\textwidth]{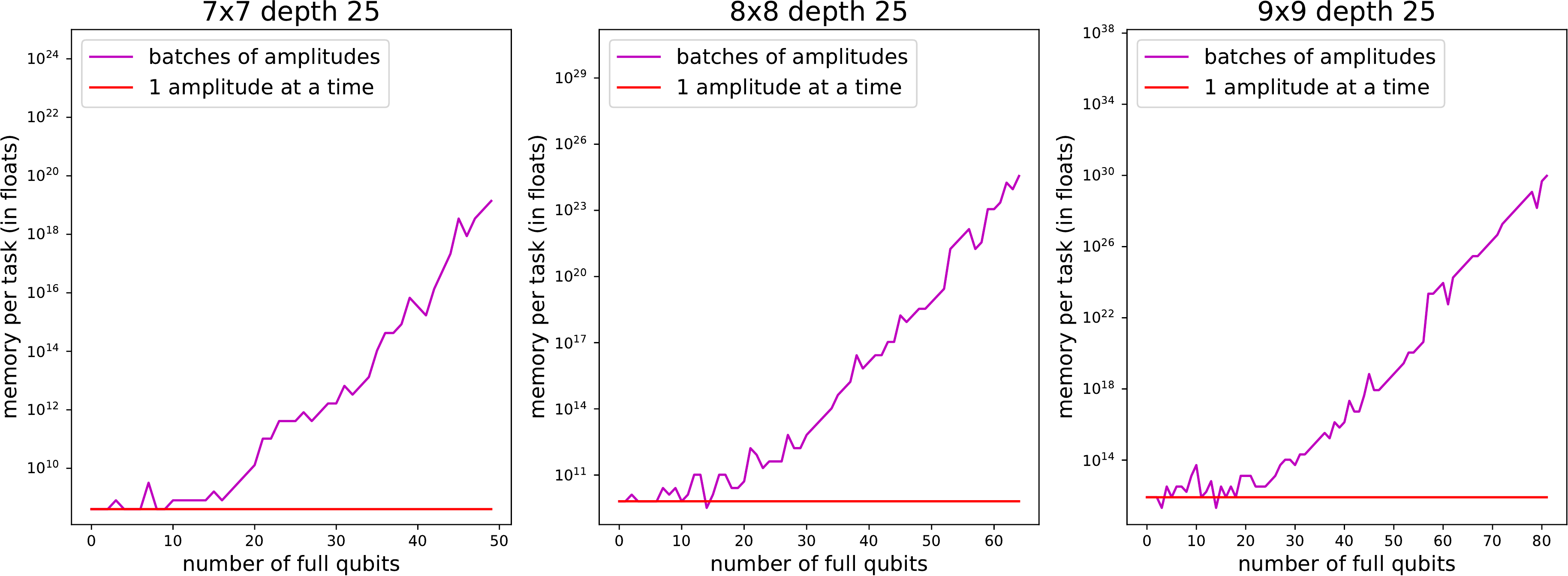}
\caption{Minimal memory requirement for the simulation of a typical random circuit of varying size}
Shown is the predicted number of memory in floating point units per single task. Notice the high similarity with Fig.~\ref{fig:flops_vs_batch_size}: the ratio of flop to memory access is almost constant in logarithmic scale. 
\label{fig:mem_vs_batch_size}
\end{figure*}

In Fig.~\ref{fig:flops_vs_batch_size}, the advantage of batch simulation comparing to single amplitude at a time is shown. The steep growth of
the flop cost is significantly ameliorated. We recall that a clique $C$ is introduced into the computational graph when the evaluation of all amplitudes of $|C|$ qubits is performed. While this clique is less than the treewidth of the computational graph, there is only a negligible increase in the computational cost. Batch simulation, however, requires copious amounts of memory, as shown in Fig.~\ref{fig:mem_vs_batch_size}. The results we obtain illustrate a usual CPU/memory trade-off seen in numerical algorithms. Notice also that the curves for total amount of memory and flop are almost indistinguishable. This result is caused by the fact that during the evaluation of the amplitudes one needs to contract high order tensors over an index of size 2: the flop cost of the most expensive contraction equals the size of the largest tensor times 2.

\begin{figure}[hbt]
\includegraphics[width=0.45\textwidth]{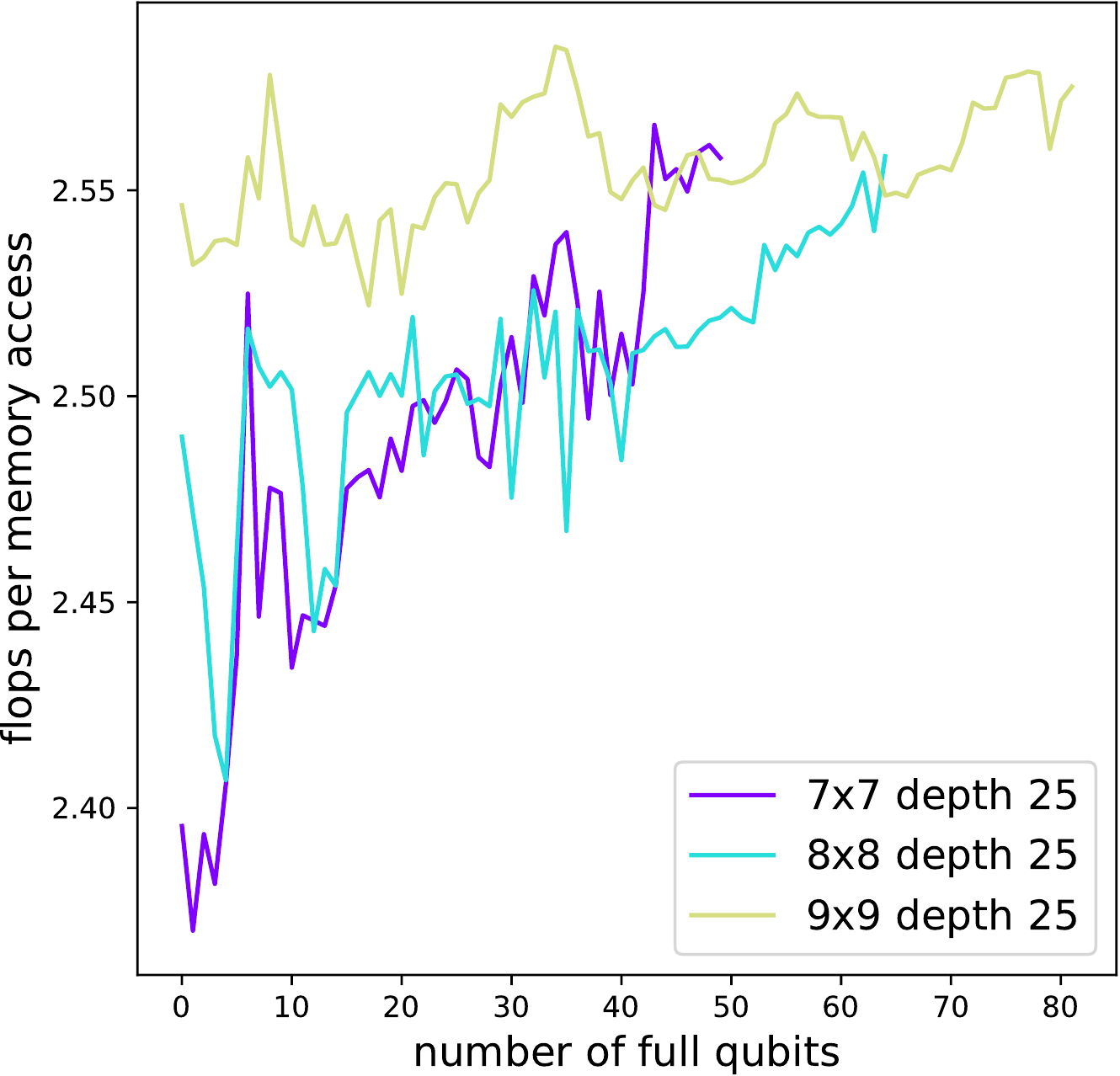}
\caption{Floating point to memory access ratio during the simulation of random circuits of varying size}
\label{fig:fpm_vs_batch_size}
\end{figure}

Lastly, we provide the dependence of the number of operations per memory access for circuits of different sizes in Fig.~\ref{fig:fpm_vs_batch_size}. In all cases, the values are in the range of $O(L)$, where $L=2$ for qubits. This dependence demonstrates that despite a potential for massive parallelism~\cite{Chen2018}, the problem of tensor contraction is essentially memory bound, and an efficient algorithm has to very carefully overlap data transmission and computations. Comparing to the contraction of matrices, where extremely efficient algorithms were developed~\cite{dongarra1988extended} using CPU cache and vectorized operations, a general tensor contraction has lower potential for optimization.

\section{Conclusion and comparison\label{sec:conclusion}}

We introduced a novel way to optimize graphical model algorithms for quantum circuits simulation. 
Our approach allows the user to select between the amount of memory consumed and the speed of the calculation; thus,
the code can be adapted to the available hardware. We emphasize that our approach is not restricted to quantum circuit simulation, but can be used to evaluate partial contractions of general tensor networks. To our best knowledge, this is a first of a kind method which evaluates partial contractions efficiently, e.g., its resource requirements depend only on the treewidth of the expression's graph.

Many more improvements to the tensor contraction
strategy can be proposed. In this article, we explicitly avoided the discussion of parallel simulation, and defer it to the upcoming publication. Also, due to the highly heterogenic structure of modern computer memories, some research is needed on the proper scheduling of the operations, especially in the parallel case.

We hope that the discussion in this article highlights a fundamental connection between tensor networks, graphs, and quantum systems.

\subsection{Acknowledgements}
The authors thank Georgy Ovchinnikov for helpful discussions. This research was conducted for the Huawei company. We kindly thank Xuecang Zhang and Yuri Zotov for their guidance, and Dmitry Kolmakov for the work on the efficient implementation of the algorithm and technical interaction with the team from Huawei's side. Patent \# XXXXXX is pending. 

\appendix
\section{Bucket elimination\label{sec:bucket_elimination}}
As was shown in the main text, graphical models are a convenient way to represent tensor contractions. One of the ways to perform contractions is the bucket elimination algorithm \cite{detcher2013bucket}. The idea of bucket elimination is simple: one starts with a graph $G = (V, E)$, which corresponds to a tensor network. Given an order $\pi: V \rightarrow \{ 0, \dots, |V| - 1 \}$, we eliminate nodes in $V$ according to $\pi$ one by one, until all nodes are removed.

The bucket elimination is implemented as follows. First buckets (sets) are formed according to the variable elimination order $\pi$. For each variable $v$ (which is also a node in $G$) we form a set of tensors $B_{\pi(v)}$ which are indexed by $v$. If both variables $v$ and $w$ index the same tensor $T$, then $T$ is placed only in the bucket corresponding to a variable with minimal order, e.g. $T \in B_{min(\pi(v), \pi(w))}$. The algorithm to form buckets is listed in \ref{code:form_buckets}.

\begin{algorithm}[H]
\caption{Forming buckets from the expression graph}\label{code:form_buckets}
\begin{algorithmic}[1]
\Require $G = (V, E), ~ G ~\mathrm{encodes ~ a ~ circuit}, \pi: V \rightarrow \mathcal{N}, ~~ \pi = \{(v_{i}, i)\}_{i=1}^{|V|}$
\Ensure $\{B_{i}\}_{i=1}^{|V|}$
\Statex
\Function{Form\_Buckets}{$G, \pi$}
\For{$i \in [1, \ldots, |V|]$}
\State $v \gets \pi^{-1}(i)$
\For{$T ~\mathrm{not ~ in} ~ \{B_{i}\}_{i=1}^{|V|}$}
\If{$T ~\mathrm{is ~ indexed ~ by} ~ v$}
\State $B_{i} \gets B_{i} \cup T$
\EndIf
\EndFor
\EndFor
\EndFunction
\end{algorithmic}
\end{algorithm}

Having arranged the tensors into the bucket structure, one can proceed to the evaluation of the expression. We process buckets according to their order. For every bucket, all tensors in it are contracted over the bucket's variable. The result is a new tensor, which is an intermediate in the contraction expression. This intermediate tensor is added to the bucket corresponding to its variable $v$ with the lowest order $\pi(v)$. The bucket processing algorithm is listed in \ref{code:process_buckets}

\begin{algorithm}[H]
\caption{Contracting expression using bucket structure}\label{code:process_buckets}
\begin{algorithmic}[1]
\Require $\mathrm{An ~ ordered ~ set ~ of ~ sets}~ \{B_{i}\}_{i=1}^{|V|} ~\mathrm{holding ~ tensors}$, $\pi = \{(v_{i}, i)\}_{i=1}^{|V|}$ 
\Ensure $\mathrm{tensor}$
\Statex
\Function{Process\_Buckets}{$B, \pi, \mathrm{stop\_index}$}
\State $\mathrm{result} ~ \gets 1$
\For{$i \in [1, \ldots, \mathrm{stop\_index}]$}
\State $v \gets \pi^{-1}(i)$
\State $T \gets \mathrm{contract ~ over} ~ v ~ \mathrm{all} ~ \tilde{T} \in B_{i}$
\If{$T ~ \mathrm{is ~ scalar}$}
\State $\mathrm{result} \gets \mathrm{result} \cdot T$
\Else
\State $k = \pi(w), ~w ~ \mathrm{indexes} ~ T, ~ w ~\mathrm{is ~ minimal ~ w.r.t.}~ \pi $
\State $B_{k} \gets B_{k} \cup T$
\EndIf
\EndFor
\State \Return result
\EndFunction
\end{algorithmic}
\end{algorithm}

It has to be noted that if one would stop the algorithm before all buckets are processed, e.g., when stop\_index $< |V|$, then the rest of the buckets $\{B\}_{\mathrm{stop\_index}}^{|V|}$ will contain a \emph{partial} contraction of the original tensor network. 

Another interesting use of bucket elimination is for the estimation of the numerical cost of tensor contractions. Indeed, instead of performing the contraction over actual tensors in the Alg.~\ref{code:process_buckets}, one can count the number of floating-point operations and the amount of used memory with their symbolic representations. Let us consider an example contraction in Eq.~\ref{eq:example_flop_count}, where for simplicity the sizes of all indices are taken to be the same (as is in the case of quantum circuits, where $L = 2$).
\begin{equation}
\begin{split}
& C_{ijk} = \sum_{l} A_{ijl} \cdot B_{jkl} \\
& dim(i) = dim(j) = dim(k) = dim(l) = L
\end{split}
\label{eq:example_flop_count}
\end{equation}

To evaluate the result of Eq.~\ref{eq:example_flop_count} one would need $L^{4}$ multiplications and additions. The whole expression would require $3 \cdot L^{3}$ of storage. The described idea can be used to estimate numerical costs of contracting a tensor network given its graphical model $G$ and an elimination order $\pi$. 

\bibliography{main}
\end{document}